\documentclass[aps,prb,twocolumn,superscriptaddress,showpacs,floatfix]{revtex4}
\usepackage{epsfig}
\usepackage[dvipsnames,usenames]{color}
\emergencystretch=\maxdimen
\begin{document}

\title{
Competing phases, phase separation and co-existence in the extended
one-dimensional bosonic Hubbard model
}

\author{G.G. Batrouni}
\affiliation{INLN, Universit\'e de Nice--Sophia Antipolis, CNRS;
1361 route des Lucioles, 06560 Valbonne, France}
\affiliation{Institut Universitaire de France, 103, Boulevard
  Saint-Michel, 75005 Paris}
\affiliation{Centre for Quantum Technologies, National University of
  Singapore, 3 Science Drive 2, Singapore 117543, Singapore} 
  
\author{V. G. Rousseau}
  \affiliation{Department of Physics and Astronomy, Louisiana State
  University, B\^aton Rouge, Louisiana 70803, USA}
  
\author{R.T.  Scalettar}
\affiliation{Physics Department, University of California, Davis,
California 95616, USA}

\author{B. Gr\'emaud}
\affiliation{Laboratoire Kastler Brossel, Ecole Normale Sup\'{e}rieure
  CNRS, UPMC; 4 Place Jussieu, 75005 Paris, France} 
\affiliation{Merlion MajuLab, CNRS-UNS-NUS-NTU International Joint
  Research Unit UMI 3654, Singapore} 
\affiliation{Centre for Quantum Technologies, National University of
  Singapore, 3 Science Drive 2, Singapore 117543, Singapore} 
\affiliation{Department of Physics, National University of Singapore,
  2 Science Drive 3, Singapore 117542, Singapore}

\begin{abstract}
  We study the phase diagram of the one-dimensional bosonic Hubbard
  model with contact ($U$) and near neighbor ($V$) interactions
  focusing on the gapped Haldane insulating (HI) phase which is
  characterized by an exotic nonlocal order parameter. The parameter
  regime ($U$, $V$ and $\mu$) where this phase exists and how it
  competes with other phases such as the supersolid (SS) phase, is
  incompletely understood. We use the Stochastic Green Function
  quantum Monte Carlo algorithm as well as the density matrix
  renormalization group to map out the phase diagram. Our main
  conclusions are that the HI exists only at $\rho=1$, the SS phase
  exists for a very wide range of parameters (including commensurate
  fillings) and displays power law decay in the one body Green
  function. In addition, we show that at fixed integer density, the
  system exhibits phase separation in the $(U,V)$ plane.
\end{abstract}

\maketitle

\section{Introduction}

The bosonic Hubbard model (BHM) has continued to attract interest
since its introduction by Fisher {\it et al.} \cite{fisher89}. This
interest stems from the versatility of the model and its use in
understanding many physical phenomena such as adsorption of bosonic
atoms on surfaces\cite{batrouni94a}, effect of disorder on superfluids
and the appearance of the compressible Bose glass
phase~\cite{fisher89}, quantum phase transitions between strongly
correlated exotic phases {\it etc}. In addition, in the hardcore
limit, the BHM can be mapped onto Heisenberg spin models and thus
offers the opportunity to study these important systems under various
conditions. Study of the BHM intensified with the experimental
realization of Bose-Einstein condensates and the ability to load them
in optical lattices~\cite{greiner02}. Under experimentally realizable
conditions, these systems are described by the BHM and its
extensions~\cite{jaksch99} with highly tunable parameters and in one,
two and three dimensions.

An increasing focus of the physics of strongly correlated quantum
systems over the last several years has been the existence of
unconventional phases and phase transitions.  In addition to well
studied Mott insulating behavior caused by an on-site repulsion, or
charge order driven by a near-neighbor repulsion, usurping
superfluidity, more exotic scenarios are realized in which different
types of order are simultaneously present, or entirely new patterns
arise.  An additional motivation for studying the extended
one-dimensional BHM is that it provides a concrete Hamiltonian in
which this physics can be examined with powerful numerical methods.

In its simplest form which has only on-site contact interactions, the
ground state of the BHM exhibits two phases~\cite{fisher89}. At
integer filling and strong repulsion, boson displacement is sterically
suppressed and the system is in an incompressible Mott insulating (MI)
phase which is replaced by a superfluid (SF) phase at weak
coupling. At incommensurate fillings, the system is always
SF. Extending this model with the addition of longer range
interactions or anisotropic hopping terms leads to new exotic
phases. For example, extensive quantum Monte Carlo (QMC) simulations
have shown that a strong enough near neighbor repulsion can lead to
insulating incompressible density wave order (CDW) at integer and half
odd integer fillings. Doping these phases can lead to phase separation
or to supersolid (SS)
phases~\cite{batrouni95,batrouni00,goral02,wessel05,boninsegni05,sengupta05,otterlo05,batrouni06,yi07,suzuki07,dang08,pollet10,capogrosso10}.

The possibility of mapping the BHM onto a Heisenberg model invites the
question of whether the same phases of the latter are present for the
former. For example, odd integer Heisenberg spin systems in
one-dimensional lattices can exhibit the exotic Haldane phase which is
a gapped phase characterized by a non-local (string) order
parameter~\cite{haldane83,dennijs89}. It was shown for the extended
one-dimensional BHM with near and next near neighbor interactions
that, at an average filling of one particle per site, the system can
be mapped approximately onto the spin-$1$ Heisenberg model and admits
a Haldane insulating (HI) phase sandwiched between MI and CDW
phases~\cite{altman06,altman08}. The phase diagram at unit filling for
the system with only contact ($U$) and near neighbor ($V$)
interactions was studied more extensively with conflicting results for
the phase diagram.  In Refs. [\onlinecite{batrouni91,batrouni94}] the
phase diagram was shown to exhibit MI, SF and CDW phases but the HI
was not found due to the very limited sizes possible to simulate at
the time. Subsequently, the $(\mu,t)$ phase diagram of the extended
BHM, for a fixed $V/U$ ratio, was obtained using Density Matrix
Renormalization Group (DMRG)~\cite{kuhner00}, but showed only evidence
for MI, SF and CDW. Recent work~\cite{rossini12}, also based on the
DMRG, has shown the presence of the HI phase between the MI and CDW
phases but found no evidence of SS at unit filling.  Even more recent
work~\cite{minguzzi} on the same model has confirmd the HI
phase. Curiously, however, there seems to be no consensus on the
nature of the phase in the ($U, V$) plane at unit filling for small
$U$ and large $V$. References
[\onlinecite{batrouni91,batrouni94,minguzzi}] show it to be SF while
Ref. [\onlinecite{rossini12}] shows it to be CDW and reference
[\onlinecite{santos}] claims it to be supersolid. We will show in this
paper that it is none of the above.

The above results give rise to some questions. Does the HI exist for
other integer fillings of the system or is it a special property of
the unit filling case? The SS phase found in one dimension
\cite{batrouni06} was obtained by doping a CDW phase: Does this phase
also exist for commensurate fillings in one dimension for parameter
choices similar to those in two \cite{kawashima12a} and three
dimensions \cite{kawashima12b}? If the SS phase exists for
commensurate fillings, where is it situated in the phase diagram
relative to the CDW, MI and HI phases?

Theoretical studies of this system using bosonization have also led to
mixed results: The HI was obtained and characterized\cite{altman08}
but consensus is absent on whether the SS phase exists in this
model. Even though older studies did not specifically mention
it~\cite{Giamarchi_book} or even argued that it did not
exist~\cite{kuhner00}, more recent studies seem to demonstrate the
presence of the SS phase ~\cite{Sengupta07,alexia}, even without
nearest neighbor interaction~\cite{Lazarides11}, for both commensurate
and incommensurate fillings. However, the precise nature of order and
the decays of the relevant correlation functions are still far from
settled.  For instance, some studies predict that the single particle
Green function decays exponentially in the SS phase while the
density-density correlation function decays as a power \cite{Lee07};
others predict that both of these correlation functions decay as
powers~\cite{Sengupta07}. Finally, the universality class of the
transition to the SS phase remains largely unexplored.

In this paper we extend our work in Ref. [\onlinecite{batrouni13}]
using the stochastic Green function (SGF) QMC algorithm~\cite{sgf} and
the density matrix renormalization group (DMRG) to study the phase
diagram of the one dimensional extended BHM as a function of the
contact (near neighbor) interaction $U$ ($V$) and the filling. For the
DMRG calculations we use the code available in the ALPS
library~\cite{alps}.  We mention that the fermionic version of this
model was also studied by means of bosonization and
DMRG\cite{barbiero} and a HI phase established.

The paper is organized as follows. In section {\bf II} we present the
model and discuss the various phases of interest and the order
parameters which characterize them. In section {\bf III} we present
our QMC and DMRG results for the phase diagrams in the $(U,V)$ plane
at fixed fillings, $\rho=1$ and $\rho=3$. We present in section {\bf
  IV} our results for the phase diagram in the $(\mu/U,t/U)$ plane at
fixed ratio $V/U=3/4$. A summary of results and conclusions is in
section {\bf V}.

\begin{table}
\begin{center}
\begin{tabular}{|l|r|r|r|r|r|r|}
\hline
× & $\rho_s$ & $S(\pi)$ & $\Delta_c$ & $\Delta_n$ & ${\cal O}_p(L_{max})$
& ${\cal O}_s(L_{max})$\\
\hline
MI & $0$ & $0$ & $\neq 0$ & $=\Delta_c$ & $\neq 0$ & $=0$\\
\hline
CDW & $0$ & $\neq 0$ & $\neq 0$ & $\neq 0$ & $\neq 0$ & $\neq 0$\\
\hline
SF & $\neq 0$ & $0$ & $0$ & $0$ & $0$ & $0$\\
\hline
HI & $0$ & $0$ & $\neq 0$ & $\neq 0$ & $0$ & $\neq 0$\\
\hline
SS & $\neq 0$ & $\neq 0$ & $0$ & $0$ & $\neq 0$ & $\neq 0$\\
\hline
\end{tabular}
\end{center}
\caption{Order parameters characterizing various phases.}
\label{phases}
\end{table}

\section{The Model}

The one dimensional extended BHM we shall study is described by the
Hamiltonian,
\begin{eqnarray}
\nonumber
 H &=& -t\sum_{i} (a^{\dagger}_ia^{\phantom\dagger}_{i+1} +
a^{\dagger}_{i+1}a^{\phantom\dagger}_{i}) + \frac{U}{2} \sum_i
n_i\left(n_i-1\right)\\
&& + V\sum_in_in_{i+1}.
\label{ham}
\end{eqnarray}
The sum over $i$ extends over the $L$ sites of the lattice, periodic
boundary conditions were used in the QMC and open conditions with the
DMRG. The hopping parameter, $t$, is put equal to unity and sets the
energy scale, $a_i^{\phantom\dagger}$ ($a_i^\dagger$) destroys
(creates) a boson on site $i$, $n_i= a_i^\dagger
a_i^{\phantom\dagger}$ is the number operator on site $i$, $U$ and $V$
are the onsite and near neighbor interaction parameters.

Several quantities are needed to characterize the phase diagram. It
was shown recently\cite{val} that the well-known expression of the
superfluid density as a function of the fluctuations of the winding
number\cite{pollock} is valid only for Hamiltonians that satisfy
\begin{equation}
[R,H] = \frac{i\hbar}{m} P,
\label{valeq}
\end{equation}
where $R$ and $P$ are the position and momentum operators and $m$ is
the mass of a boson. Here $t=\hbar^2/(2m {\ell}^2)$ where $\ell$ is the
lattice constant. It is straightforward to verify that Eq.(\ref{ham})
satisfies this condition and, therefore, the superfluid density is
given by\cite{pollock}
\begin{equation}
\rho_s = \frac{\langle W^2\rangle}{2td\beta L^{d-2}},
\label{rhos}
\end{equation}
where $W$ is the winding number of the boson world lines, $d$ is the
dimensionality and $\beta$ the inverse temperature. The CDW order
parameter is the structure factor, $S(k)$, at $k=\pi$ where
\begin{equation}
S(k) = \frac{1}{L}\sum_{r=0}^{L-1} {\rm e}^{ikr}\langle n_0n_r
\rangle,
\label{sk}
\end{equation}
and the momentum distribution, $n_k$, is given by
\begin{equation}
n_k=\frac{1}{L}\sum_{r=0}^{L-1} {\rm e}^{ikr}\langle a^{\dagger}_0
a^{\phantom\dagger}_r \rangle.
\label{nk}
\end{equation}
The charge gap is given by,
\begin{eqnarray}
\label{chargegapmu}
\Delta_c(n)&=&\mu(n)-\mu(n-1)\\
\nonumber
&=& E_0(n+1)+E_0(n-1)-2E_0(n)
\label{chargegapE}
\end{eqnarray}
where $\mu(n)=E_0(n+1)-E_0(n)$ and $E_0(n)$ is the ground state energy
of the system with $n$ particles and is obtained both with QMC and
DMRG. The neutral gap, $\Delta_n$, is obtained using DMRG by targeting
the lowest excitation with the same number of bosons. In both CDW and
HI phases, the chemical potentials at both ends are set to (opposite)
large enough values, in DMRG, such that the ground state degeneracy
and the low energy edge excitations are
lifted~\cite{kuhner00,altman06}.  With the SGF we did simulations in
both the canonical and grand canonical ensembles.
\begin{figure}[h!]
\epsfig{figure=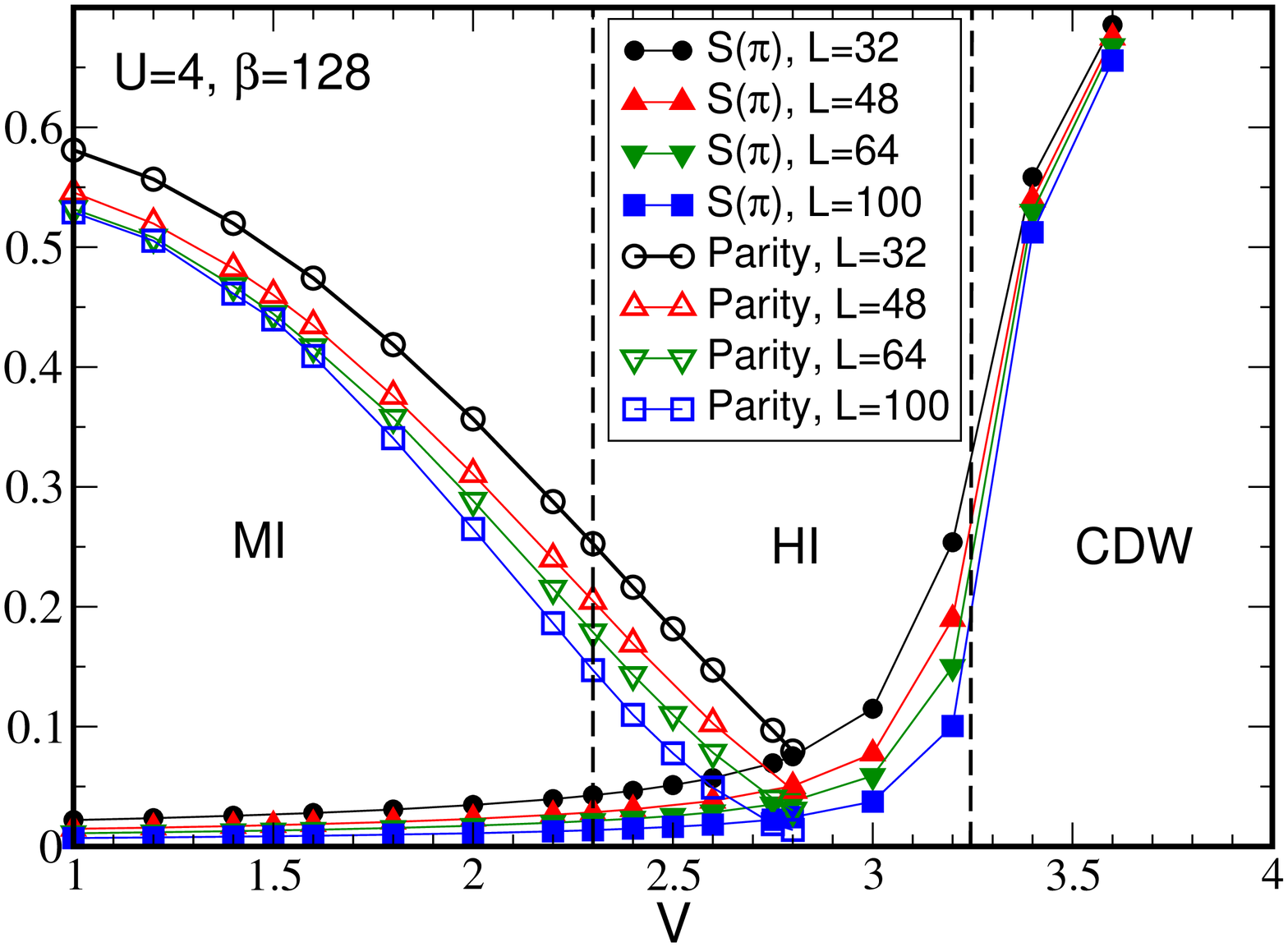,width=9.5cm,clip}\\
\vskip -0.5cm
\epsfig{figure=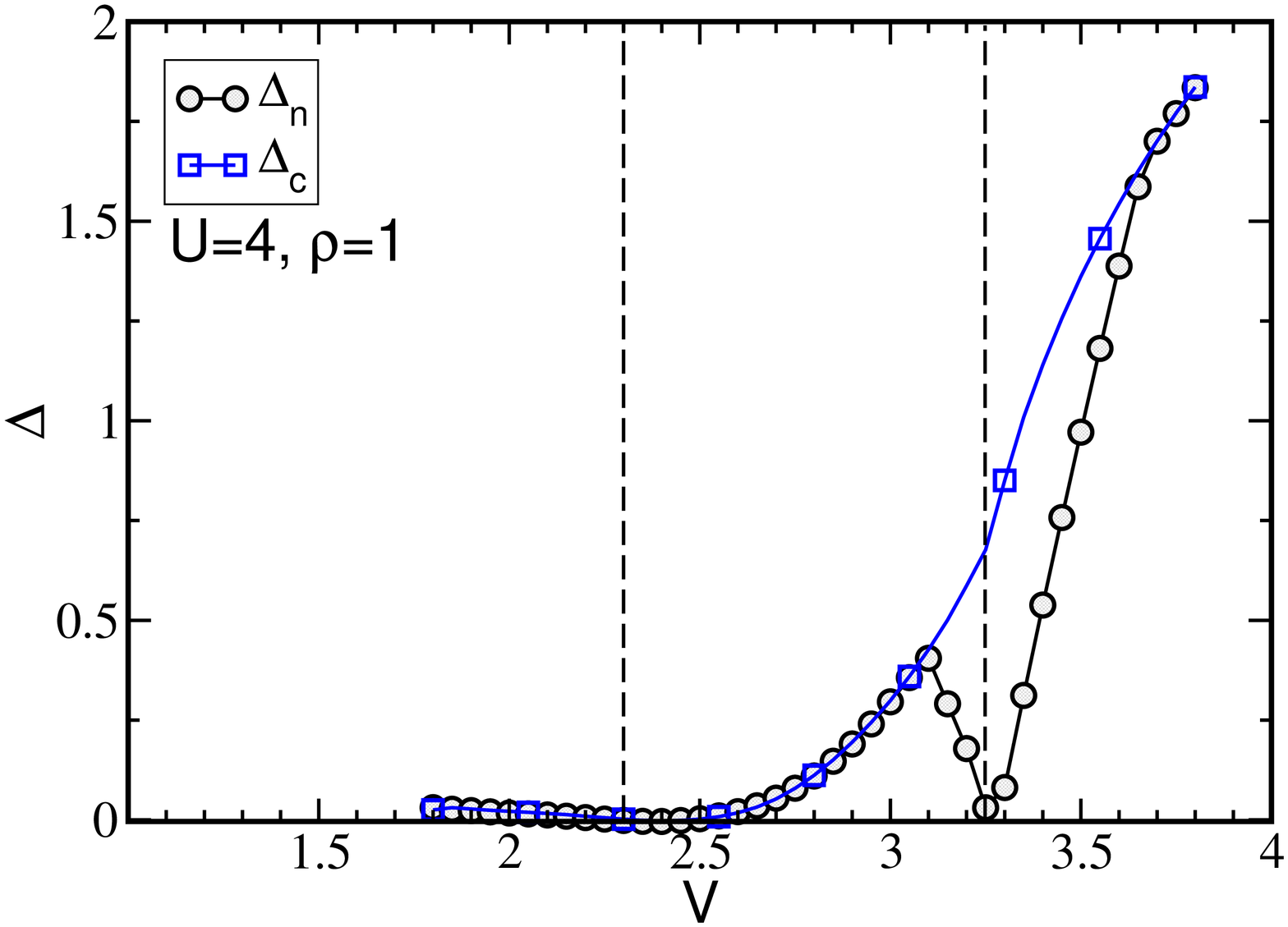,width=9.5cm,clip}
\caption{(color online) Top: The parity, ${\cal O}_p$, and CDW,
  $S(\pi)$, order parameters versus $V$ at fixed $U=4$ and several
  system sizes using QMC.  By extrapolating the order parameters,
  $L\to \infty$, we identify two critical values, $V_1^c=2.3$, and
  $V_2^c=3.25$. For $V< V_1^c$, the system is in the $\rho=1$ MI; for
  $V> V_2^c$, the system is in the CDW phase. For $V_1^c<V<V_2^c$
  (between the two vertical dashed lines), both ${\cal O}_p$ and
  $S(\pi)$ vanish as $L\to \infty$ but the string order parameter,
  ${\cal O}_s$, (not shown) remains finite. This is the HI
  phase. NOTE: ${\cal O}_p$ has been multiplied by $5$ for all sizes
  to render curves more visible in the figure. Bottom: The
  extrapolated charge, $\Delta_c$, and neutral, $\Delta_n$, gaps for
  the same system using DMRG. Both gaps vanish at the MI-HI transition
  but only the neutral gap vanishes at the HI-CDW transition, as
  expected. The critical values confirm the QMC results in the top
  panel.}
\label{rho1constU}
\end{figure}

For spin systems, the nonlocal Haldane string order parameter is given
by,
\begin{eqnarray}
{\cal O}_s(|i-j|\to \infty) &=& \langle S^z_i {\rm exp}(i\theta
  \sum_{k=i}^j S^z_k) S^z_j \rangle
\label{stringspin}
\end{eqnarray}
where $\theta=\pi$ for the spin-$1$ system. This order parameter
detects the Haldane phase where the $S_z=1$ and $S_z=-1$ alternate
along the lattice and are separated by varying numbers of $S_z=0$
sites. In other words, the system exhibits, in this phase, long range
anti-ferromagnetic order but with no characteristic momentum. In the
BHM at $\rho=1$, when $U$ and $V$ are large (with $2V<U$), most sites
are singly occupied. Quantum fluctuations allow some sites to be
unoccupied or doubly occupied, but higher occupations are
suppressed. One can therefore make the analogy with spin systems and
define $S_z(i)\equiv \delta n_i=n_i-\rho$ ($\rho=1$) and define two
nonlocal order parameters for the BHM, the string and the parity
parameters:
\begin{eqnarray}
{\cal O}_s(|i-j|\to \infty) &=& \langle \delta n_i {\rm e}^{i\theta \sum_{k=i}^j
  \delta n_k} \delta n_j \rangle,\\
\label{string}
{\cal O}_p(|i-j|\to \infty) &=& \langle  {\rm e}^{i\theta \sum_{k=i}^j \delta
  n_k} \rangle.
\label{parity}
\end{eqnarray}
In practice we take the order parameters to be ${\cal
  O}_{s/p}(L_{max})$ where, in QMC with PBC, $L_{max}=L/2$ and in
DMRG, with OBC, $L_{max}$ is the longest distance possible before edge
effects start being felt.  For higher integer filling,
$\rho=2,3\dots$, $\theta\neq \pi$ and has to be determined as
discussed in Ref. [\onlinecite{qin03}].
\begin{figure}[h!]
\centerline{\epsfig{figure=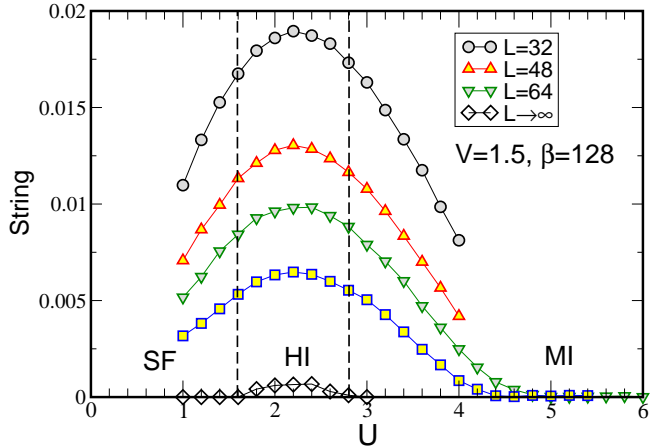,width=9.5cm,clip}}
\caption{(color online) The string order parameter, ${\cal O}_s$,
  versus $U$ for $V=1.5$ and several sizes. Also shown are the
  extrapolated values giving the two critical values $U_1^c=1.6$ and
  $U_2^c=2.8$. For $U_1^c\leq U \leq U_2^c$ (the interval between the
  vertical dashed lines) the system is in the HI. For $U< U_1^c$, the
  system is in the SF phase and for $U>U_2^c$, it is in the MI
  phase. }
\label{stringvsU}
\end{figure}

The various thermodynamically stable phases that may appear in the
system are characterized by the above quantities and are summarized in
Table \ref{phases}. To explore the possibility of phase separation
(PS), it is not enough to look at order parameters since if the system
is in a phase separated state (a mixture of two or more thermodynamic
phases), there will be contributions from the order parameters of all
the phases present. To preclude or confirm the presence of phase
separation, we study the behavior of the density $\rho$ as a function
of the chemical potential, $\mu$, and also the density profile in the
system.

\begin{figure}[h!]
\centerline{\epsfig{figure=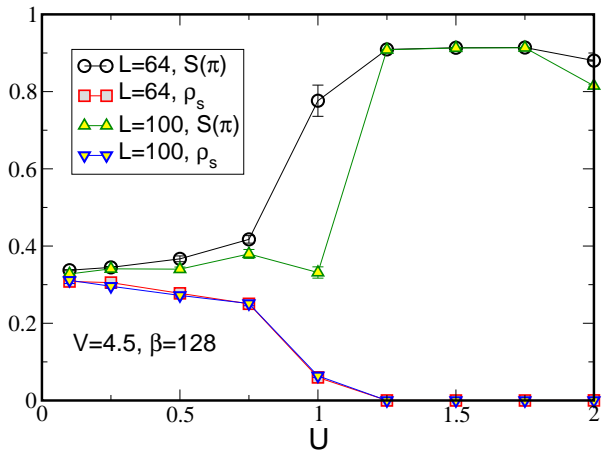,width=9.5cm,clip}}
\caption{(color online) The superfluid density, $\rho_s$, and CDW
  order parameter $S(\pi)$ as functions of $U$ at fixed $V=4.5$ and
  $\rho=1$. For $U>1$: $\rho_s=0$ and $S(\pi)\neq 0$ indicate a CDW
  phase. For $U\lesssim 1$: $\rho_s\neq 0$ and $S(\pi)\neq 0$ suggest
  the possibility of a supersolid phase.}
\label{rhosSpiV4.5}
\end{figure}

\section{Phase diagrams at fixed fillings}

We start with the phase diagram in the ($U,V$) plane at fixed unit
density. To map out the phase diagrams, we fix the filling at a
commensurate value (here we focus on $\rho=1$ and $\rho=3$) and we fix
one of the interaction parameters, $U$ or $V$, while the other is
varied. The physical quantities discussed above ($\rho_s$, $S(k=\pi)$,
${\cal O}_p$, ${\cal O}_s$) are calculated and the various phases
deduced from Table \ref{phases}. For example, in the top panel of
Fig.~\ref{rho1constU} we show ${\cal O}_p$ and $S(\pi)$ as functions
of $V$ for the system at $\rho=1$ and fixed $U=4$ and several lattice
sizes. Extrapolating these quantities to $L\to \infty$ yields the two
critical values $V^c_1=2.3$, where ${\cal O}_p$ vanishes, and
$V^c_2=3.25$ where $S(\pi)$ becomes nonzero. For all $V$ at $U=4$,
$\rho_s=0$ (not shown in Fig. \ref{rho1constU}) and, therefore, there
is no SF phase at $U=4$. The string order parameter, ${\cal O}_s$,
vanishes for $V<V^c_1$ (not shown in the figure) and takes on a finite
value for $V>V^c_1$. According to Table \ref{phases}, this means that
for $V<V^c_1$ the system is in the MI phase; for $V^c_1<V<V^c_2$ the
system is in the HI phase and for $V>V^c_2$, the system enters the CDW
phase.
\begin{figure}[h!]
\centerline{\epsfig{figure=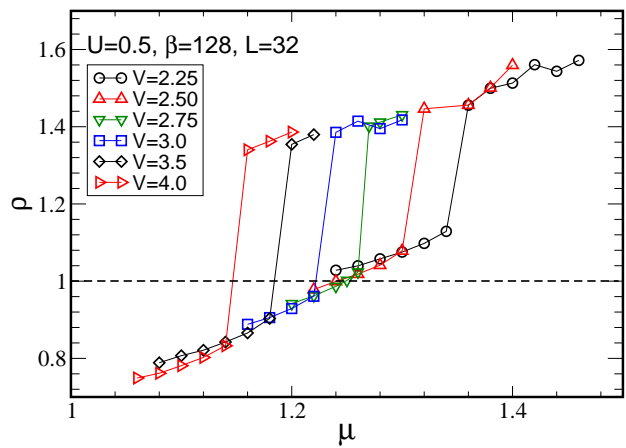,width=9.5cm,clip}}
\caption{(color online) The density, $\rho$, as a function of the
  chemical potential, $\mu$, for several values of the near neighbor
  interaction, $V$, at fixed contact interaction, $U=1/2$. For all
  values of $V$ there is a discontinuous jump in $\rho(\mu)$
  indicating a first order phase transition. For $V \geq 3$, the
  $\rho=1$ value (indicated by the dashed line) is inside the
  jump. This means that there is no thermodynamically stable phase
  with $\rho=1$ for these values of $V$. This figure was obtained with
  QMC simulations in the grand canonical ensemble.}
\label{rhovsmu}
\end{figure}

These transitions can also be seen in the behavior of $\Delta_c$ and
$\Delta_n$, the charge and neutral gaps, shown in the lower panel of
Fig. \ref{rho1constU} as functions of $V$ at $\rho=1$ and $U=4$. These
are the extrapolated gap values, i.e., $L\rightarrow\infty$, from the
DMRG results obtained for sizes ranging from $L=64$ to $L=256$.  We
see that at the MI-HI transition, $V=2.3$, both gaps vanish and at the
HI-CDW transition, $V=3.25$, $\Delta_n$ vanishes while $\Delta_c$ does
not. This behavior is expected\cite{altman08} and the DMRG values
agree with those obtained via QMC (top panel Fig. \ref{rho1constU}).

The behavior of the string order parameter is shown in
Fig. \ref{stringvsU} as a function of $U$ for fixed $V=1.5$ and for
several sizes. Also shown are the extrapolated values using ${\cal
  O}(L)={\cal O}(L\to \infty) +{\rm const./L}$ and demonstrating that
for $1.6<U<2.8$, ${\cal O}_s$, although small, is
nonvanishing. Therefore, the system is in the HI in this interval. For
$U<1.6$ the system is SF since $\rho_s \neq 0$. For $U>2.8$, upon
examining the other quantities such as ${\cal O}_p$ (nonzero) and
$\rho_s$ (vanishes) we conclude that the system is in the MI phase.

\begin{figure}[h!]
\centerline{\epsfig{figure=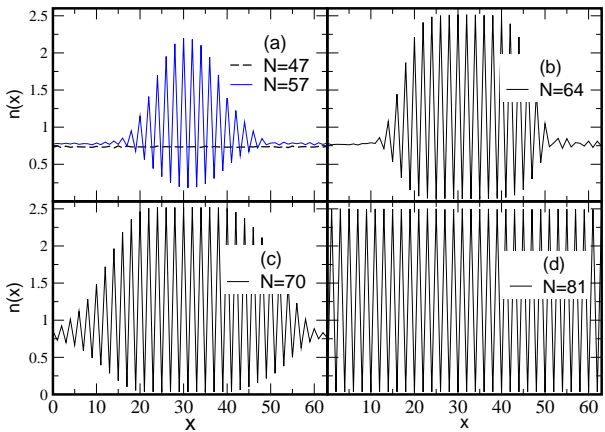,width=9.5cm,clip}}
\caption{(color online) When, in a canonical simulation, the number of
  particles is fixed at a value which corresponds to a
  thermodynamically unstable phase, the system undergoes phase
  separation. In this figure, we show QMC results for $L=64$, $U=1/2$
  and $V=4.5$. The phase with $N=47$ particles, dashed horizontal line
  in (a), is seen to be uniform and corresponds to the SF
  phase. Adding a few particles, however, destabilizes the system as
  when there are $N=57$ particles. As more particles are added, the
  size of the region displaying CDW increases as is seen in (b) for
  $N=64$ and (c) for $N=70$. In (d), with $N=81$ the system appears to
  be entirely CDW. However, this CDW region is in fact a
  supersolid. One way to see that is to notice that the density
  oscillates between $0$ and $2.5$. Also, the system in (d) exhibits
  both long range density order (CDW) and non-vanishing superfluid
  density. Note: In the QMC simulation, the CDW region does not always
  appear in the same place. Since these simulations are done with
  periodic boundary conditions, we have centered these regions to
  facilitate comparison.}
\label{denprof}
\end{figure}

The behavior of the system in the region at small $U$ and large $V$ is
shown in Fig. \ref{rhosSpiV4.5} for $V=4.5$. For $U> 1$, $\rho_s=0$
and $S(\pi)\neq 0$ signalling the presence of CDW. On the other hand,
for $U \lesssim 1$, both $\rho_s$ and $S(\pi)$ are nonzero. This
simultaneous finiteness of $S(\pi)$ and $\rho_s$ seems to indicate
that the system is in the supersolid phase. To confirm this, however,
one must show that this is a thermodynaimcally stable phase and not a
mixture of two phases. To this end, we show in Fig. \ref{rhovsmu} the
density, $\rho$, as a function of the chemical potential, $\mu$, for
several values of $V$ and fixed $U=1/2$. All these curves, which were
obtained by QMC in the grand canonical ensemble, exhibit discontinuous
jumps in the density at critical values of $\mu$ which depend on
$V$. This shows that, for these values of $U$ and $V$, the system
exhibits a first order phase transition. Furthermore, it is clear from
Fig. \ref{rhovsmu} that for $V\geq 3$ (and $U=1/2$) the discontinuous
jump in $\rho$ includes the value $\rho=1$. This means that if the
number of particles is fixed at $\rho=1$, {\it i.e.} if one considers
the canonical ensemble, then for $U=1/2$ and $V\geq 3$, the system
undergoes phase separation.

\begin{figure}[h!]
\centerline{\epsfig{figure=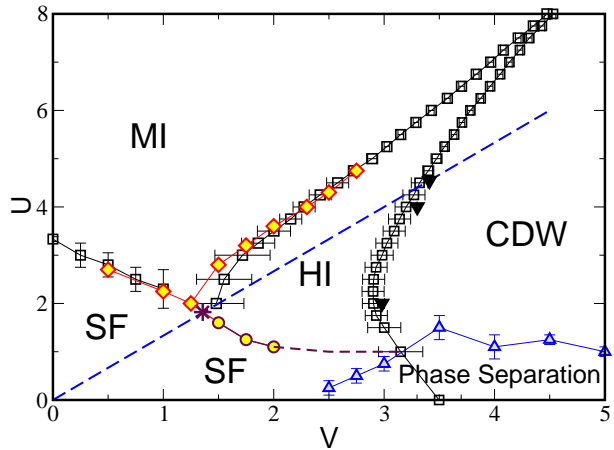,width=9.5cm,clip}}
\caption{(color online) The phase diagram in the ($U,V$) plane for
  $\rho=1$. The open squares are the results of
  Ref. \onlinecite{rossini12}, all other symbols are results of our
  QMC simulations. Our results confirm those of
  Ref. \onlinecite{rossini12} where the error bars are small and
  improve them where the error bars are large. In addition, we have
  determined the SF-HI boundary and also the boundary of the phase
  separation region.  The dashed line is given by $U=4V/3$ and will be
  discussed in the text.}
\label{phasediagrho1}
\end{figure}

By studying $\rho(\mu)$ in this way for various values of $U$ and $V$
we map out the region in the ($U,V$) plane where phase separation
takes place. The question then is: What are the phases into which the
system separates? To answer this, we examine the density profiles (in
the canonical ensemble) in the system at various fillings. In
Fig. \ref{denprof} we show the density profiles for five fillings,
$N=47,\, 57,\, 64,\, 70,\, 80$ on a lattice with $L=64$. For $N=47$
(dashed line in Fig \ref{denprof}(a)) the density is uniform and the
system is SF. As more particles are added, for example for $N=57$ in
Fig. \ref{denprof}(a), the system becomes inhomogeneous developing two
regions, one with uniform density (and still SF) and one with
alternating site densities giving the appearance of a developing CDW
phase. As more particles are added, the region with alternating site
densities expands at the expense of the uniform one,
Fig. \ref{denprof}(c)(d). When enough particles are added, the system
becomes uniform displaying an oscillating local density profile
indicative of a CDW phase. However, the site density alternates
between $0$ and $2.5$ indicating that it is, in fact, a supersolid
(SS) phase because in a true CDW phase, the local density will
alternate between $0$ and an integer value. This is confirmed by
measurement of $\rho_s$. Therefore, the observed phase separation is
between the SF and SS phases. Note that in the QMC simulation, the CDW
region does not always appear in the same place. Since these
simulations are done with periodic boundary conditions, we have
centered these regions to facilitate comparison. In addition, results
obtained with the DMRG exhibit similar density profiles and give rise,
therefore, to the same boundaries for the phase separation in the
$(U,V)$ plane. From a mean-field point of view, the usual Gutzwiller
ansatz~\cite{Iskin} only predicts a second order SF-SS phase
transition and, therefore, no phase separation for $\rho=1$. Finally,
since the occupation number in the density profiles shown in
Fig.~\ref{denprof} ranges from 0 to almost 2.5, a mapping of the
bosonic Hubbard model to a spin chain model would require values of
the total spin $S$ larger than 1, which explains that this phase
separation is not present in the $S=1$ chain.  Actually, for a fixed
value of $V$, one can clearly see that $S$ increases with decreasing
values of $U$: for $U=0$, the CDW can have an arbitrary number of
bosons on every other site.  It is therefore tantalizing to map this
situation to the classical anisotropic Heisenberg model with a single
ion anisotropy, which does have a first order phase transition as a
function of the magnetic field ($\mu$ for bosons), the spin-flop
transition. However, this transition is between the Neel order and the
spin-flop phase, which, in the bosonic language, corresponds to a
transition between the CDW and the SF phases, not between the SS and
the SF ones. A proper mean-field understanding of the present
transition is thus still lacking.
\begin{figure}[h!]
\centerline{\epsfig{figure=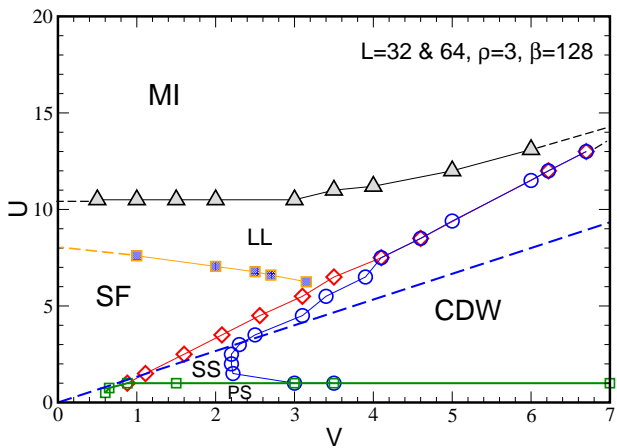,width=9.5cm,clip}}
\caption{(color online) Applying the same methods as for the $\rho=1$
  case, the $\rho=3$ phase diagram is mapped out. Unlike for $\rho=1$,
  there is no HI here but there is SS. The region of the Luttinger
  liquid phase (LL) with the Luttinger parameter $K>2$ is indicated by
SF and where $K<2$ by LL}
\label{phasediagrho3}
\end{figure}

The resulting phase diagram at fixed $\rho=1$ is shown in
Fig. \ref{phasediagrho1}. The open squares are the results of
Ref. [\onlinecite{rossini12}], all other symbols are results of our
QMC simulations. Our results confirm some of the results of
Ref. \onlinecite{rossini12}, in particular where their error bars are
small, and improve them where their error bars are large. In addition,
we have determined the SF-HI boundary and also show the newly found
phase separation region. The dashed line is given by $U=4V/3$ and will
be discussed below.

Applying the same techniques at fixed filling $\rho=3$ gives the phase
diagram Fig. \ref{phasediagrho3}. As for the case of $\rho=1$,
Fig. \ref{phasediagrho1}, the $\rho=3$ phase diagram,
Fig. \ref{phasediagrho3}, exhibits MI, SF, CDW phases and phase
separation. However, it does not exhibit the HI phase. Instead,
sandwiched deep between the CDW and MI phases, is a Luttinger liquid
(LL) phase with $\langle a^{\phantom\dagger}_0 a^{\dagger}_r\rangle
\sim r^{-1/2K}$ and parameter $K<2$ and is, therefore, not SF. In
addition, the $\rho=3$ phase diagram exhibits a supersolid phase not
present at $\rho=1$. The absence of the HI at $\rho=3$ is likely due
to the fact that here, unlike for $\rho=1$, the bosonic system cannot
be simply mapped on to a Heisenberg spin chain system.

\begin{figure}[h!]
\centerline{\epsfig{figure=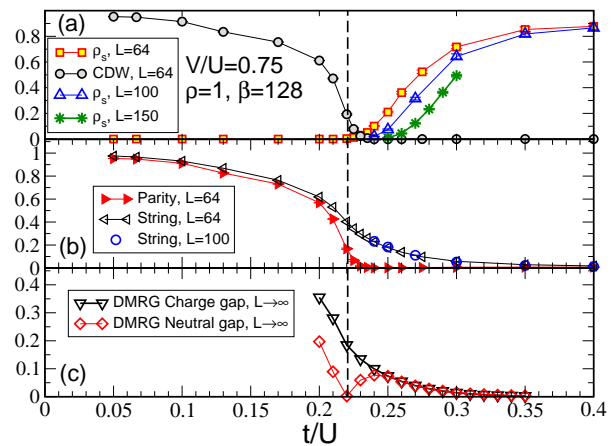,width=9.5cm,clip}}
\caption{(color online) Shows several quantities for $\rho=1$ as
  functions of $t/U$ with the fixed ratio $V/U=0.75$. (a) The CDW
  order parameter for $L=64$ and $\rho_s$ for $L=64,\, 100,\, 150$;
  (b) the parity and string order parameters; (c) the neutral and
  charge gaps. (a) and (b) were obtained with QMC and (c) with
  DMRG. The region to the right of the dashed line is the HI. It
  terminates at $t/U=0.55$. In the HI, $\rho_s\to 0$ very slowly as
  $L$ increases. Note the difference between the neutral and charge
  gaps. The gaps are given in units of the hopping $t$.}
\label{rho1orders}
\end{figure}

\section{Phase diagram at fixed $V/U=3/4$}

In this section we study the phase diagram in the ($\mu/U,t/U$) plane
at fixed ratio $V/U=3/4$. This value of $V/U$ is chosen because the
CDW phase is favored over the MI phase at large $U$ and integer
filling\cite{batrouni13}.

\begin{figure}[h!]
\centerline{\epsfig{figure=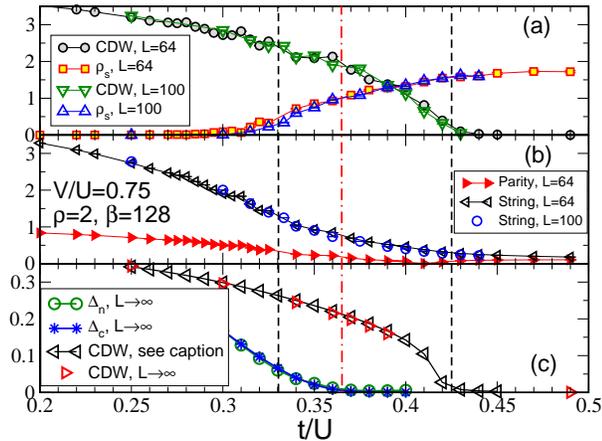,width=9.5cm,clip}}
\caption{(color online) Same as for $\rho=1$ but at $\rho=2$. (a) QMC
  simulations show that in the interval between the two vertical
  (black) dashed lines there is simultaneous SF and CDW order and,
  therefore, a supersolid phase (SS). The vertical (red) dot-dash line
  is where the $L\to \infty$ extrapolated neutral ($\Delta_n$) and
  charge ($\Delta_c$) gaps vanish in DMRG (c). The CDW-SS transition
  is between $t/U=0.33$ (QMC) and $t/U=0.355$ (DMRG). The difference
  between the two values could be due to the difference in the
  boundary conditions, open for DMRG and periodic for QMC. (c) also
  shows the $L\to \infty$ extrapolated CDW order parameter, right
  (red) triangles, and the Fourier transform of $\langle
  n_i\rangle\langle n_j\rangle $, left (black) triangles, obtained
  with DMRG to probe the disappearance of CDW order. Both DMRG and QMC
  give the SS-SF transition at $t/U\approx 0.425$.  Note that, unlike
  Fig.~\ref{rho1orders}, the charge and neutral gaps (c) are essentially
  always the same. }
\label{rho2orders}
\end{figure}

We start by fixing $\rho=1$ and studying the various phases as $t/U$
is changed (with $V/U=3/4$). This path is shown as the dashed straight
line in Figs. \ref{phasediagrho1} ($\rho=1$) and \ref{phasediagrho3}
($\rho=3$). Figure \ref{rho1orders} shows the behavior of $\rho_s$,
$S(k=\pi)$, ${\cal O}_s$, ${\cal O}_p$, $\Delta_c$ and $\Delta_n$ as
$t/U$ is changed. For $t/U < 0.23$, $\rho_s=0$ and $S(\pi)\neq 0$
(Fig. \ref{rho1orders}(a)) indicating that the system is in the CDW
phase. Also in the CDW phase ($t/U<0.23$) ${\cal O}_p\neq 0$ and
${\cal O}_s\neq 0$ and are both essentially equal to the CDW order
parameter, $S(\pi)$. As the transition out of the CDW phase is
approached, $t/U \to 0.23$, ${\cal O}_p\to 0$ as does
$S(\pi)$. However, ${\cal O}_s$ decreases but remains finite
indicating that the phase is HI for $t/U>0.23$. This scenario is
confirmed in Fig. \ref{rho1orders}(c) which shows the neutral and
charge gaps, $\Delta_n$ and $\Delta_c$. In this panel we see that in
the CDW phase, $\Delta_c > \Delta_n$ and $\Delta_n\to 0$ as $t/U\to
0.23$ while $\Delta_c$ remains nonzero. In other words, the neutral
gap vanishes at the CDW-HI transition but the charge gap remains
nonzero showing the HI phase to be gapped. As $t/U$ is increased
further, the system eventually transitions into the SF phase which is
indicated by the star symbol on the dashed straight line in
Fig. \ref{phasediagrho1}.
\begin{figure}[h!]
\centerline{\epsfig{figure=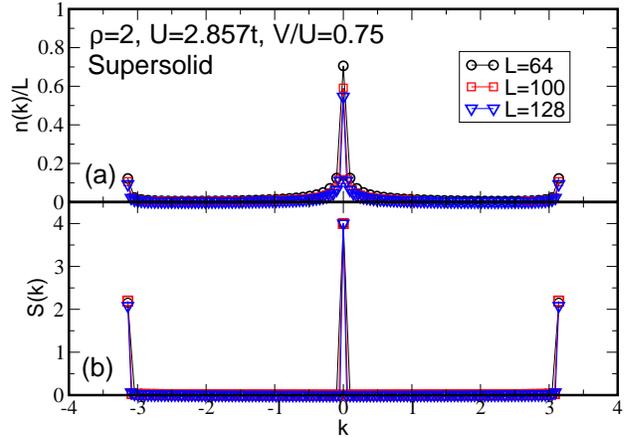,width=9.5cm,clip}}
\caption{(color online) The dependence of the momentum distribution,
  $n_k/L$, and the structure factor, $S(k)$, on the system
  size. $n_k/L \to 0$ with increasing $L$ while $S(k)$ remains
  constant indicating long range CDW order.  }
\label{SSnk}
\end{figure}

As mentioned previously, the behavior at $\rho=1$ may be understood by
drawing on the analogy with the $S=1$ spin system. The question then
arises as to whether the same analogy holds for the other integer
fillings. To answer this question we perform the same analysis but for
$\rho=2$. The results are shown in Fig. \ref{rho2orders}. Figure
\ref{rho2orders}(a) shows that for $t/U>0.33$, $\rho_s$ increases from
zero but $S(\pi)$ remains nonzero. $S(\pi)$ vanishes for $t/U > 0.42$
(confirmed by DMRG in panel (c) of the same figure). In other words,
unlike the $\rho=1$ case, there is an interval where the system
exhibits simultaneous long range diagonal (density) order and
superfluidity. This is the hallmark of the SS phase. In the $\rho=1$
case, the HI intervenes between the CDW and SF phases while at
$\rho=2$ the SS phase takes that role. To confirm the nature of the SS
phase, we show in Fig. \ref{SSnk} the structure factor, $S(k)$ and the
momentum distribution, $n(k)$, for several lattice sizes. The fact
that the peak $S(\pi)$ does not change with the system size indicates
that true long range order in the density is present. The fact that
the peak at $n(k=0)$ decreases as $L$ increases is expected since
there cannot be true off diagonal long range order in one dimension.

\begin{figure}[h!]
\centerline{\epsfig{figure=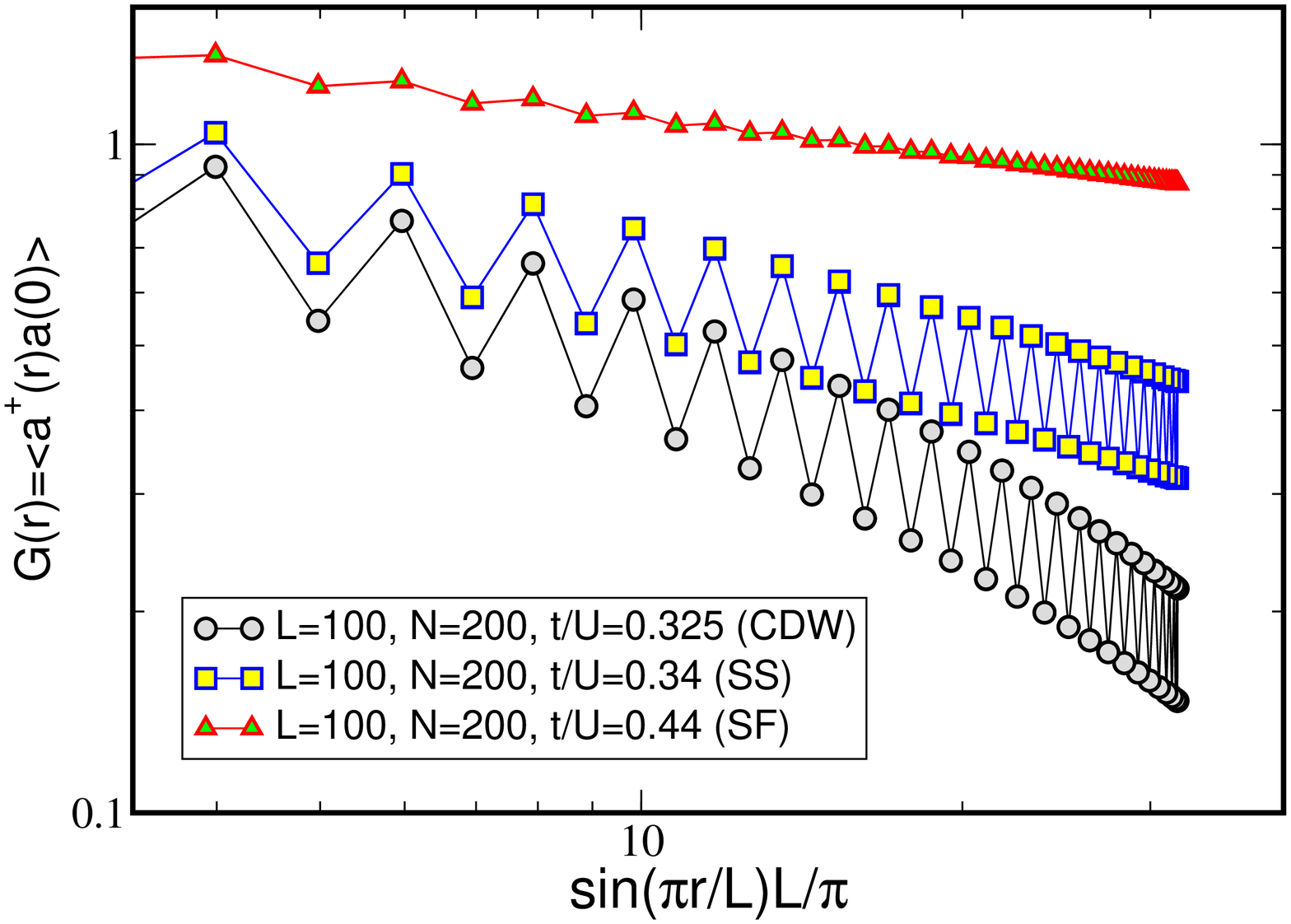,width=9.5cm,clip}}
\caption{(color online) The single particle Green function decays as a
  power law in the SF ($K=2.5$) and SS ($K=1.2$) phases and exponentially
  in the CDW phase. In the SS phase, $G(r)$ also also exhibits
  oscillations due to the long range density order.}
\label{greenrho2}
\end{figure}

The behavior of the single particle Green function in the SS phase is
clarified in Fig. \ref{greenrho2} which shows $G(x)$ for $\rho=2$ in
the SF, SS and CDW phases.  The QMC is done with periodic boundary
conditions and consequently, the Green function will be symmetric with
respect to $r=L/2$. To handle this, we plot in the figure, on semi-log
scale, the Green function versus $x=(L/\pi){\rm sin}(\pi r/L)$ whose
limit as $L\to \infty$ is $x=r$. It is seen that in the SF and SS
phases, $G(x)$ decays as a power ($K=2.5$ and $K=1.2$ respectively)
although in the SS phase there are modulations due to the long range
density order. In the CDW phase, $G(x)$ decays exponentially as
expected.

The same behavior is observed for $\rho=3$ and, in fact, for
$\rho=5/2$. For the $\rho=3/2$ case, it appears that the system exits
the CDW phase directly into the SF phase (see below). 

\begin{figure}[h!]
\centerline{\epsfig{figure=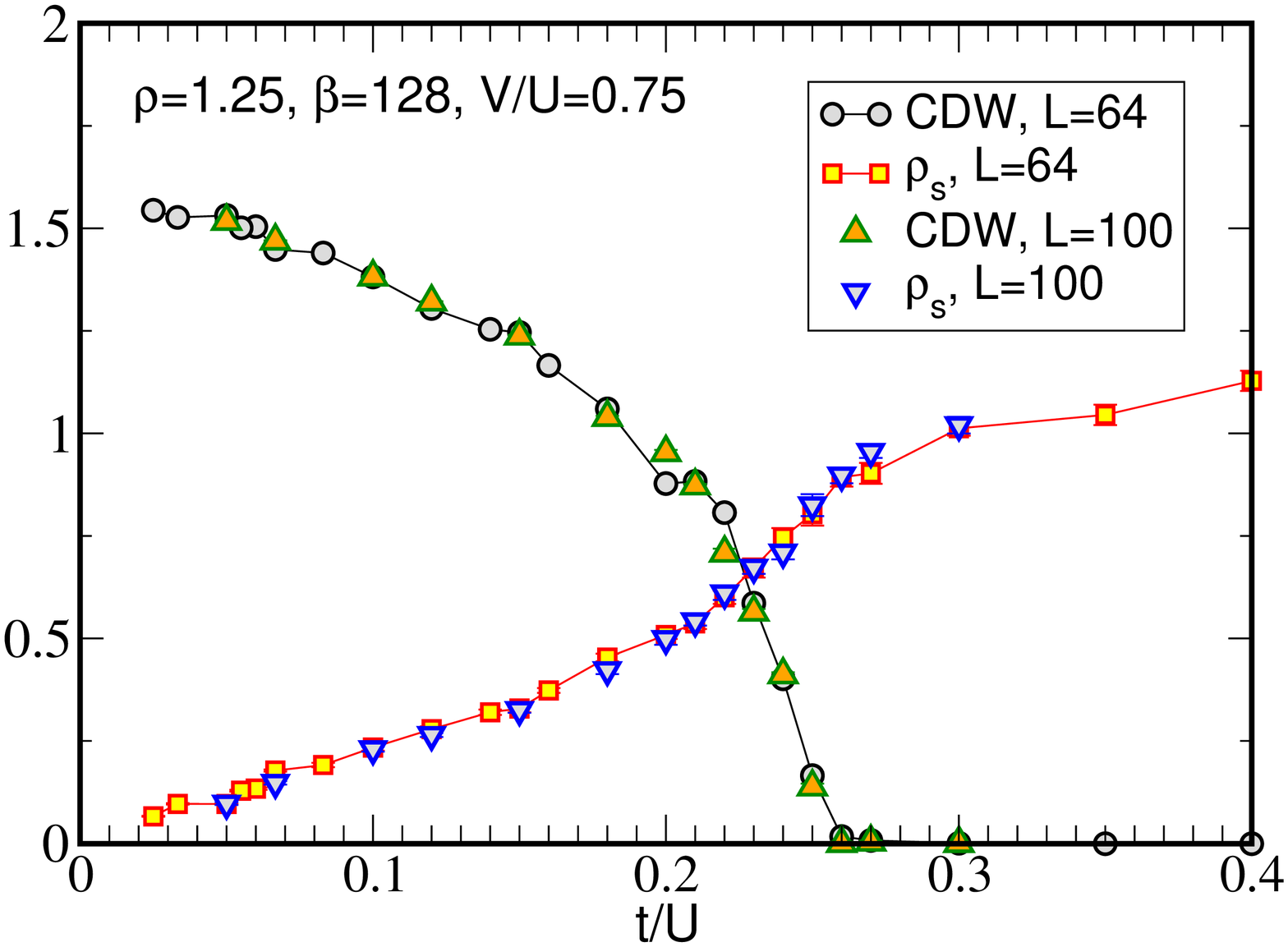,width=9.5cm,clip}}
\centerline{\epsfig{figure=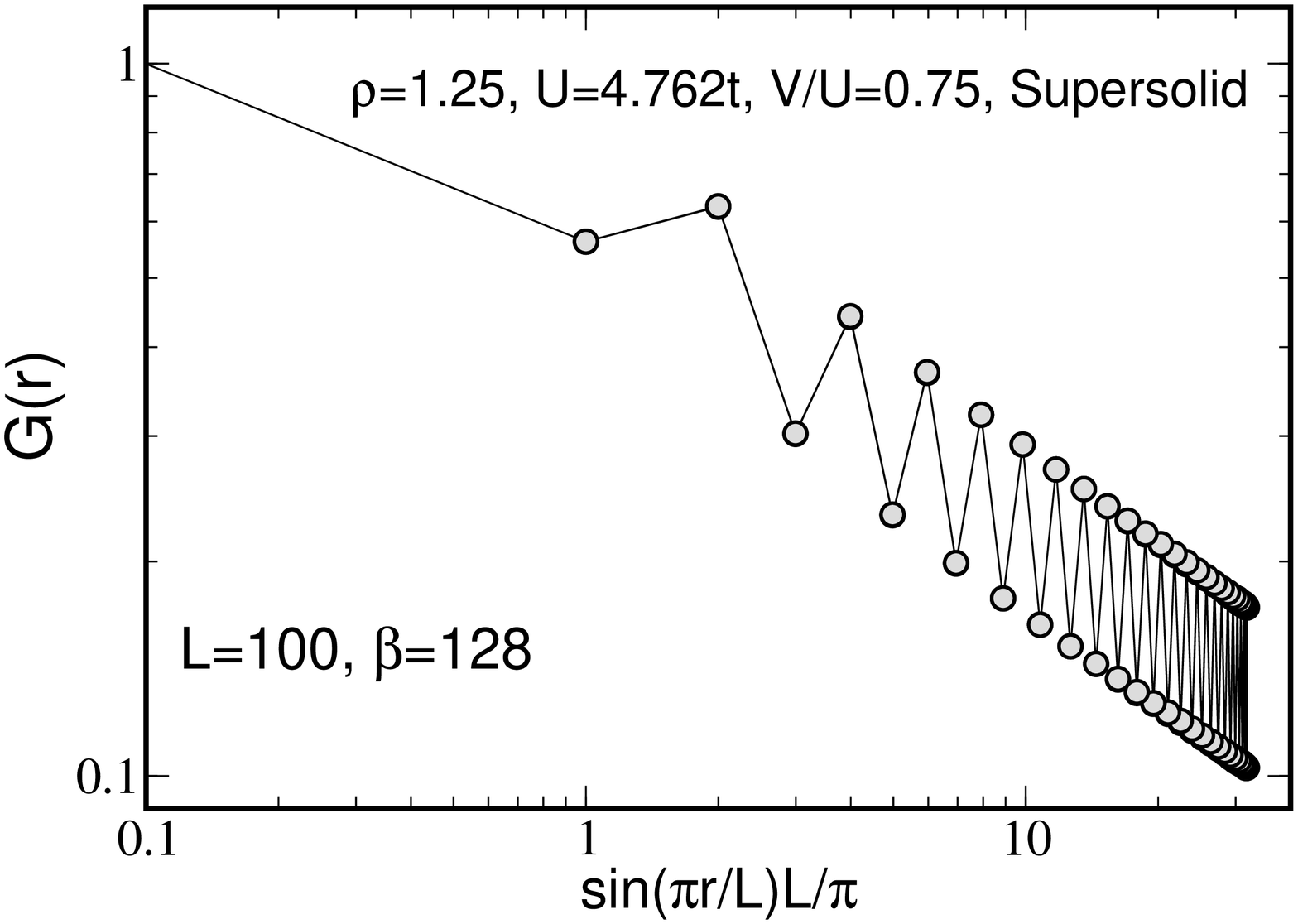,width=9.3cm,clip}}
\caption{(color online) Top: The superfluid density and the CDW order
  parameter as functions of $t/U$ at $\rho=1.25$. The system passes
  from the supersolid phase to the superfluid phase as $t/U$ is
  increased. Bottom: The one-particle $G(r)$ decays as a power in the
  SS phase with $K\approx 1.1$.}
\label{rho1.25}
\end{figure}

To map out the phase diagram in the ($\mu/U, t/U$) plane we need to
characterize the phases at incommensurate fillings. In the top panel
of Fig. \ref{rho1.25}, we show $\rho_s$ and $S(\pi)$ as functions of
$t/U$ at a filling of $\rho=1.25$ and $V/U=3/4$. We see that two
phases are present: SS, where CDW and SF are present simultaneously,
and SF.

\begin{figure}[h!]
\centerline{\epsfig{figure=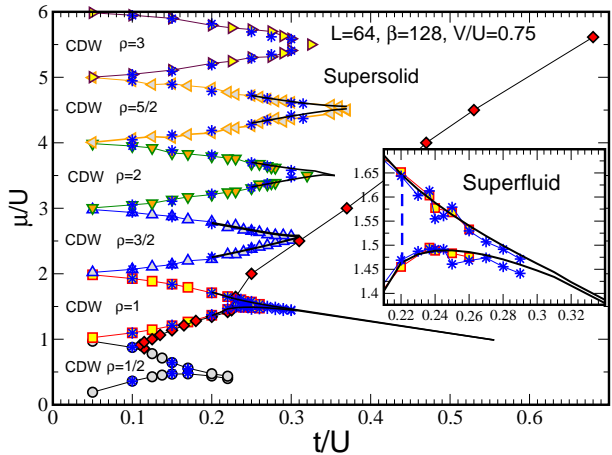,width=9.3cm,clip}}
\caption{(color online) The phase diagram at fixed ratio
  $V/U=3/4$. The inset is a zoom on the tip of the $\rho=1$ lobe. All
  symbols represent results from QMC simulations for $L = 128$ (stars)
  and $L = 64$ (all other symbols).  The solid black lines near the
  lobe tips are DMRG results with $L = 192$. The end points of the
  lobes are obtained by studying the finite size dependence of
  $\Delta_n$ using DMRG except for $\rho=1$ which is obtained using
  QMC by extrapolating ${\cal O}_s$ to the thermodynamic limit.}
\label{phasediag}
\end{figure}

So, for fixed $V/U=3/4$, we have shown the presence of four phases:
CDW, HI, SS and SF. By performing scans such as those that led to
Figs. \ref{rho1orders}, \ref{rho2orders}, \ref{rho1.25} and by
calculating the charge gaps for the CDW phase, we find the boundaries
of these phases and map the phase diagram in the ($\mu/U,t/U$)
plane. The phase diagram is shown in Fig. \ref{phasediag}. In
Fig. \ref{phasediag}, all symbols represent results from QMC
simulations for $L = 128$ (stars) and $L = 64$ (all other symbols),
$\beta = 128$.  The solid black lines near the lobe tips are obtained
from DMRG with $L = 192$. The end points of the lobes are obtained by
studying the finite size dependence of $\Delta_n$ using DMRG except
for $\rho=1$ which is obtained using QMC by extrapolating ${\cal O}_s$
to the thermodynamic limit (the star symbol in
Fig. \ref{phasediagrho1}). The inset is a zoom of the tip of the $\rho
= 1$ lobe.

\begin{figure}[h!]
\centerline{\epsfig{figure=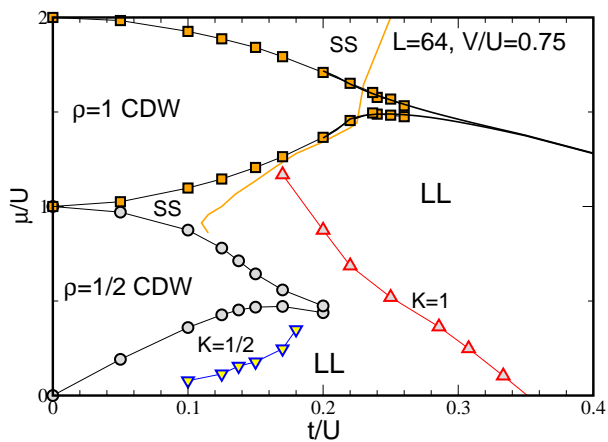,width=9.3cm,clip}}
\caption{(color online) Detail of the $\rho=1/2$ lobe where we also
  determined the constant $K$ lines for $K=1,\, 1/2$. The Luttinger
  liquid in this region ($K<2$) is not SF. }
\label{rhohalflobe}
\end{figure}

Several comments are in order. Except for a small region of SS
squeezed between it and the $\rho=1$ lobe, the $\rho=1/2$ lobe is
surrounded almost entirely by LL phase with $K<2$ and, therefore, not
SF, see Fig. \ref{rhohalflobe}. The fact that in the extended BHM a SS
does not exist when the $\rho=1/2$ CDW phase is doped with holes, but
does when it is doped with particles, was already addressed in
Ref. [\onlinecite{batrouni06}]. The $\rho=1$ lobe sticks out of the SS
phase and the part sticking out is, in fact, the HI phase. No other
CDW lobe behaves this way. The $\rho=3/2$ lobe terminates right at the
boundary with the SF phase: To within the resolution of our
simulations, the transition from the $\rho=3/2$ CDW lobe goes directly
into the SF phase without passing through the SS phase.  This peculiar
behavior for $\rho=3/2$ was also observed with additional DMRG results
for different values of $V/U$ ranging from $0.65$ to $1$: The SS layer
between the CDW and SF phases, if present, is too thin to observe for
the considered system sizes. An accurate determination of the $(U,V)$
phase diagram for this filling will require a more thorough finite
size scaling analysis.  All other CDW lobes, $\rho\geq 2$, are
surrounded entirely by the SS phase. It is interesting to compare this
figure with Fig.~3 of Ref. [\onlinecite{kawashima12a}] and with the
mean-field predictions~\cite{Iskin}.

\section{Conclusions}

Even though the one dimensional BHM with near neighbor interaction is
a rather simple model, it continues to attract attention and to yield
surprises such as new phases with exotic order parameters and quantum
phase transition.

In this paper we used the SGF QMC algorithm and DMRG to elaborate the
details of the phase diagram of the extended BHM in one dimension and
expose novel features and exotic phases. We mapped the phase diagrams
in the $(\mu/U,t/U)$ plane at fixed $V/U=3/4$ and in the $(U,V)$ plane
at two fixed commensurate fillings, $\rho=1,3$. We find that, for this
system, the HI seems to exist only at $\rho=1$, invalidating the
Heisenberg spin analogy at higher integer fillings. We study the
charge and neutral gaps and the nonlocal string order parameter
characterizing this phase. For higher densities, we find that the
supersolid phase, SS, is very robust and exists for a very wide range
of parameters including at commensurate fillings. We show that the
one-body Green function decays as a power in the SS phase, not
exponentially as sometimes argued. We also showed that when the
filling is fixed, there exists a region in the $(U,V)$ plane where the
system undergoes phase separation. This phase separated region can be
mistaken for a supersolid phase if only the order parameters $\rho_s$
and $S(\pi)$ are studied. Evaluation of $\rho(\mu)$ and also the
spatial density profile reveals the phase separation unambiguously.

\acknowledgments We thank T. Giamarchi for very helpful
discussions. This work was supported by: the CNRS-UC Davis EPOCAL
joint research grant; by the France-Singapore Merlion program (PHC
Egide and FermiCold 2.01.09); by the LIA FSQL; by grant DOE
DE-NA0001842-0. The Centre for Quantum Technologies is a Research
Centre of Excellence funded by the Ministry of Education and National
Research Foundation of Singapore.


\end{document}